\documentclass[12pt,a4paper,twoside]{article}
\usepackage[dvips]{graphicx}
\usepackage{cite}
\usepackage{layout}

\newcommand{\bbbar}  {\ensuremath{\mathrm{b\overline{b}}}}
\newcommand{\epem}   {\ensuremath{\mathrm{e^+e^-}}}

\newcommand{\msb}    {\ensuremath{\mathrm{\overline{MS}}}}
\newcommand{\lmqcd}  {\ensuremath{\Lambda_{\mathrm{QCD}}}}
\newcommand{\lmqcdsq}{\ensuremath{\Lambda^2_{\mathrm{QCD}}}}
\newcommand{\as}     {\ensuremath{\alpha_s}}
\newcommand{\ash}    {\ensuremath{\hat{\alpha}_s}}
\newcommand{\asmu}   {\ensuremath{\alpha_s(\mu)}}

\newcommand{\asq}    {\ensuremath{\alpha_s(Q)}}
\newcommand{\asmz}   {\ensuremath{\alpha_s(M_{\mathrm{Z^0}})}}
\newcommand{\oa}     {\ensuremath{\mathcal{O}(\alpha_s)}}
\newcommand{\oaa}    {\ensuremath{\mathcal{O}(\alpha_s^2)}}
\newcommand{\oaaa}   {\ensuremath{\mathcal{O}(\alpha_s^3)}}
\newcommand{\znull}  {\ensuremath{\mathrm{Z^0}}}

\newcommand{\mz}     {\ensuremath{M_{\mathrm{Z^0}}}}

\newcommand{\bt}     {\ensuremath{B_T}}
\newcommand{\bw}     {\ensuremath{B_W}}

\newcommand{\thr}    {\ensuremath{1-T}}
\newcommand{\cp}     {\ensuremath{C}}
\newcommand{\chisq}  {\ensuremath{\chi^2}}
\newcommand{\chisqd} {\ensuremath{\chi^2/\mathrm{d.o.f.}}}
\newcommand{\xmu}    {\ensuremath{x_{\mu}}}

\newcommand{\fit}    {\ensuremath{\mathrm{(fit)}}}
\newcommand{\scale}  {\ensuremath{\mathrm{(scale)}}}
\newcommand{\had}    {\ensuremath{\mathrm{(had.)}}}
\newcommand{\exptl}  {\ensuremath{\mathrm{(exp.)}}}

\newcommand{\bm}[1]  {\mbox{\boldmath\ensuremath{#1}}}
\newcommand{\lnr}    {\ensuremath{\ln(R)}}

\newcommand{\roots}  {\ensuremath{\sqrt{s}}}

\newcommand{\cf}     {\ensuremath{C_F}}
\newcommand{\ca}     {\ensuremath{C_A}}
\newcommand{\tf}     {\ensuremath{T_F}}
\newcommand{\nf}     {\ensuremath{n_f}}
\newcommand{\mui}    {\ensuremath{\mu_I}}
\newcommand{\mil}    {\ensuremath{{\cal M}}}
\newcommand{\epsb}   {\ensuremath{\epsilon_b}}

\newcommand{\anull}   {\ensuremath{\alpha_0}}

\newcommand{\stan}    {fit result}
\newcommand{\fite}    {fit error}
\newcommand{\rensc}   {ren. scale}
\newcommand{\hadr}    {hadr.}
\newcommand{\tote}    {tot. error}

\begin{document}

%
%

\begin{titlepage}

\bigskip\bigskip

\begin{flushright}
       MPI-PhE/2000-19   \\ 
       Revised version \\
       1 June 2001 \\
\end{flushright}

\bigskip\bigskip\bigskip\bigskip\bigskip

\begin{center}
{\huge\bf
A Measurement of the QCD Colour Factors using Event Shape
Distributions at \bm{ \roots=14 } to 189 GeV 
}
\end{center}

\bigskip\bigskip

\begin{center}
{\Large S.~Kluth, P.~A.~Movilla Fern\'andez, S.~Bethke,
C.~Pahl and P. Pfeifenschneider } \\
\bigskip
{\large
Max-Planck-Institut f\"ur Physik, \\
Werner-Heisenberg-Institut \\
F\"ohringer Ring 6 \\
80805 Munich, Germany \\ }
\end{center}

\bigskip\bigskip\bigskip

\begin{abstract}
Measurements of the QCD colour factors \ca\ and \cf\ and of the number
of active quark flavours \nf\ in the process
$\epem\rightarrow\mathrm{hadrons}$ at high energy are presented. They
are based on fits of \oaa+NLLA QCD calculations to distributions of
the event shape observables \thr, \cp, \bt\ and \bw\ measured at
centre-of-mass energies from 14~GeV to 189~GeV.  Hadronisation effects
are approximated with power correction calculations which also depend
on the QCD gauge structure. In this approach potential biases from
hadronisation models are reduced. Our results for individually
measured quantities obtained from \thr\ and \cp\ are
\begin{displaymath}
\nf = 5.64\pm1.35,\;\;\;
\ca = 2.88\pm0.27\;\;\;\mathrm{and}\;\;\;
\cf = 1.45\pm0.27 
\end{displaymath}
in good agreement with QCD based on the SU(3) symmetry group where
$\nf=5$ for the energies considered here, $\ca=3$ and $\cf=4/3$. From
simultaneous fits of \ca\ and \cf\ with \thr\ and \cp\ we find
\begin{displaymath}
\ca = 2.84\pm0.24\;\;\;\mathrm{and}\;\;\;
\cf = 1.29\pm0.18\;,
\end{displaymath}
which is also in good agreement with the QCD expectation. 
\end{abstract}

\bigskip\bigskip\bigskip\bigskip

%
\begin{center}
{\large 
(Submitted to European Physical Journal C) }
\end{center}

\end{titlepage}

\section{ Introduction }

The theory of strong interactions, Quantum Chromo Dynamics (QCD),
gives a successful description of most aspects of hadronic final
states in \epem\  annihilation and other high energy particle
collision processes over a large range
of centre-of-mass (cms) energies, see e.g.
\cite{OPALPR299,OPALPR303,OPALPR197,OPALPR158,jadenewas,jadec,h1jetqcd,zeusf2,d0jetsqcd,biebel01a}.
In particular, the strong coupling \as, the only free parameter in the
theory, can be measured to an accuracy of about
3\%~\cite{bethke00a}. 

Standard QCD contains three distinct colour charges which are carried
by the quarks and obey the SU(3) symmetry group, see
e.g.~\cite{muta87,ellis96}. Requiring local gauge invariance of the
theory generates eight massless vector fields, the gluons, each
carrying a colour charge and an anticolour charge. Since the gluons
are colour charged QCD has more fundamental vertices than QED, where
the gauge boson, the photon, is electrically neutral. The fundamental
vertices are i) the radiation of a gluon from a quark line (gluon
bremsstrahlung), ii) the conversion of a gluon into a quark-antiquark
pair (quark pair production), iii) the conversion of a gluon into two
gluons (triple gluon vertex) and iv) the conversion of a gluon into
three gluons (quartic gluon vertex). The quartic gluon vertex is in
\epem\ annihilation of order \oaaa\ and thus inaccessible to the \oaa\
analysis presented here (see section~\ref{sec_pred}).

The relative weights of each of the vertices are given by the QCD
colour factors \cf, \tf\ and \ca\ corresponding to vertices i), ii)
and iii), respectively~\cite{muta87}. The colour factor \tf\ for vertex
ii) contributes for each of \nf\ active quark flavours, such that the
product $\tf\nf$ is the weight for vertex ii).

The colour factors can be calculated from the generators of gauge
transformations in the fundamental and adjoint representation once a
particular symmetry group has been chosen. In standard QCD the
symmetry group is SU(3) and the corresponding values of the colour
factors are $\cf=4/3$, $\tf=1/2$ and $\ca=3$. Perturbative QCD
calculations, e.g. for cross sections, depend on the symmetry group
only through the colour factors. It is thus straightforward to modify
the prediction for a different symmetry group by recalculating the
colour factors.

From an experimental point of view one can ask the question if SU(3)
is indeed the correct symmetry group for QCD. This question may be
tested experimentally by allowing the colour factors to vary in
comparisons of QCD predictions with data. Several such studies have
been performed by the LEP experiments based on correlations of angles
in 4-jet final
states~\cite{angcorrl3,OPALPR019,4jetdelphi1,4jetaleph,4jetdelphi2,OPALPR113,4jetdelphi3}.
In these studies the colour factors enter at leading order (\oaa) of
the QCD calculations and thus higher order corrections have not been
taken into account. Measurements of the ratios $\nf/\cf$ and $\ca/\cf$
by this method have been in agreement with the QCD expectations.

A complementary approach has been to employ distributions of event
shape observables measured at LEP~1 and
\oaa+NLLA\footnote{NLLA stands for Next-to-Leading-Log Approximation,
see section~\ref{sec_pred}.} QCD
predictions\cite{magnoli90,OPALPR134}. Here the sensitivity to the colour
factors also enters at \oaa\ but higher order corrections are
partially taken care of by the inclusion of the NLLA terms.
More recent works combined the two analysis techniques outlined
above~\cite{aleph4jet2} and also made use of recently calculated
higher order corrections to the 4-jet angular
correlations~\cite{dienes00}. In both cases results have been
consistent with QCD predictions and systematic uncertainties
were reduced\footnote{The error treatment of the results
of~\cite{aleph4jet2} have been criticised~\cite{farrar97}.}. 

It is also possible to use the dependence of QCD predictions on the
energy scale of the hard process to gain sensitivity to the colour
factors~\cite{hebbeker93,l3lep2data1,bethke00a}. The dependence of QCD
predictions on the energy scale enters through the running of the
strong coupling \as\ which in turn is a function of the colour
factors.

A common property of all studies discussed so far is that they rely on
Monte Carlo models of the hadronisation process, typically
JETSET\cite{jetset3}, HERWIG~\cite{herwig} or ARIADNE~\cite{ariadne3}. 
Dependence on these hadronisation models in studies of the QCD gauge
structure may be viewed as a disadvantage, because the models assume
standard QCD with the SU(3) symmetry group to be valid. No attempt was
made to study the effects of non-standard colour factors on
the hadronisation corrections. However, such effects could be
significant, because the parton shower stages in the Monte Carlo models
are based on QCD calculations which in turn depend on the QCD gauge
symmetry.

Recently an analytic model of the hadronisation process has become
available~\cite{dokshitzer95a} commonly referred to as power
corrections. It is based on the analysis of the power-behaviour of
non-perturbative effects, i.e. the growth of non-perturbative effects
with inverse powers of the scale of the hard process. In this model
approximate predictions for the non-perturbative effects on mean
values and distributions of some event shape observables have been
made and their colour structure is explicitly
given~\cite{dokshitzer95,dokshitzer97a,dokshitzer98a,nllacp,dokshitzer99a,dokshitzer99b}.
Comparison with data has been reasonably
successful~\cite{jadenewas,jadec,delphilep2data1,l3lep2data1,h1powc,movilla00a}.

In this study we use fits of \oaa+NLLA QCD predictions with power
corrections to differential distributions of the event shape
observables Thrust, C-parameter, Total and Wide Jet Broadening
measured in \epem\ annihilation at cms energies from 14~GeV to 189~GeV
to investigate the gauge structure of QCD. Section~\ref{sec_pred}
introduces the observables and summarises the QCD
predictions. Section~\ref{sec_fits} gives a description of the data
sets used in the analysis and of the fit results. Finally in
section~\ref{sec_summ} a summary and conclusions are given.

\section{ QCD Predictions }
\label{sec_pred}

\subsection{ Event Shape Observables }

We use the following event shape observables:
\begin{description}
\item[Thrust \bm{T}] This observable is defined by the
  expression~\cite{thrust1,thrust2} 
  \begin{equation}
    T= \max_{\vec{n}}\left(\frac{\sum_i|\vec{p}_i\cdot\vec{n}|}
                    {\sum_i|\vec{p}_i|}\right)
    \label{equ_thrust} 
  \end{equation}
  where $p_i$ is the momentum of reconstructed particle $i$ in an event. 
  The thrust axis $\vec{n}_T$ is the direction $\vec{n}$ for which the
  maximum occurs. We will use the form \thr\ here since its
  distribution is in this form more similar to those of the other
  observables.  
\item[C-parameter \bm{\cp}] The definition of this
  observable~\cite{parisi78,donoghue79} requires the introduction of
  the linearised energy-momentum tensor  
  \begin{equation}
    \Theta^{\alpha\beta} = \frac{ \sum_i(p_i^{\alpha}p_i^{\beta})/|\vec{p}_i| }
                                { \sum_i|\vec{p}_i| } \;\;\;, \;\;\; 
    \alpha,\beta=1,2,3
  \end{equation}
  and its three eigenvalues $\lambda_k$, $k=1,2,3$. These define \cp\
  via 
  \begin{equation}
  \cp= 3(\lambda_1\lambda_2+\lambda_2\lambda_3+\lambda_3\lambda_1)\;\;\;.
  \end{equation}
  Note that \cp\ may equivalently be calculated by
  $\cp=(3/\sum_i|p_i|)\sum_{i<j}|\vec{p}_i||\vec{p}_j|\sin^2(\theta_{ij})$
   where $\theta_{ij}$ is the angle between particles $i$ and $j$. 
\item[Jet Broadening] The definitions of the jet broadening
  observables~\cite{nllabtbw} employ a plane through the origin
  perpendicular to the thrust axis $\vec{n}_T$ to divide the event
  into two hemispheres $S_1$ and $S_2$. The Total and the Wide Jet
  Broadening \bt\ and \bw\ are defined as
  \begin{equation}
    B_i= \frac{ \sum_{p\in S_i} |\vec{p}\times\vec{n}_T| }
              { 2\sum_j |\vec{p}_j| }\;\;\;, \;\;\; 
    \bt= B_1+B_2\;\;\;, \;\;\;
    \bw= \max(B_1,B_2)\;\;\;.
  \end{equation}
\end{description}
For these observables complete \oaa+NLLA QCD predictions as well as
power correction calculations for their differential distributions are
available.

\subsection{ Running of \as }

The running of the strong coupling \as\ is a direct consequence of the
requirement that a complete QCD calculation, e.g. for a cross section
$R$, should not depend on the choice of energy scale where the theory
has been renormalised, formally expressed as the renormalisation group
equation (RGE), see e.g.~\cite{ellis96}:
\begin{equation}
\mu^2\frac{dR}{d\mu^2} = 
\mu^2\left( \frac{ \partial }{ \partial\mu^2 } +
\frac{\partial\as}{\partial\mu^2}\frac{\partial}{\partial\as}
\right) R = 0\;\;\;.
\label{equ_rge1}
\end{equation}
The running of \as\ is found by solving the following differential
equation where $\beta(\as(\mu^2))$ is the beta function of QCD shown
at two-loop accuracy~\cite{ellis96}:
\begin{equation}
\mu^2\frac{\partial\as}{\partial\mu^2} = \beta(\asmu) =
-\beta_0\as^2(\mu) - \beta_1\as^3(\mu)\;\;\;,
\label{equ_rge2}
\end{equation}
\begin{displaymath}
\beta_0=\frac{11\ca-2\nf}{12\pi} \;\;\;\mathrm{and}\;\;\;
\beta_1=\frac{17\ca^2-5\ca\nf-3\cf\nf}{24\pi^2}\;\;\;.
\end{displaymath}
The coefficients $\beta_0$ and $\beta_1$ are independent of the
renormalisation scheme. The solution to equation~(\ref{equ_rge2}) at
this order is~\cite{OPALPR134,ellis96}
\begin{equation}
\beta_0\ln(\xmu^2)= \frac{1}{\asmu} - \frac{1}{\asq} +
\frac{\beta_1}{\beta_0}\ln\left(\frac{\asmu}{\asq}\cdot
\frac{\beta_0+\beta_1\asq}{\beta_0+\beta_1\asmu}\right)
\label{equ_runas}
\end{equation}
with $\xmu=\mu/Q$. The quantity \xmu\ is commonly referred to as the
renormalisation scale factor since it allows to study the effects of
variations of the renormalisation scale in perturbative QCD
predictions. It is usually varied to assess the effect of missing
higher order in the perturbation series.  Equation~(\ref{equ_runas})
may be solved numerically to obtain \asmu\ as a function of a
reference \asq. The QCD gauge structure enters via the coefficients
$\beta_0$ and $\beta_1$ of the QCD $\beta$-function.

The perturbative evolution of \as\ diverges for small scales, see
e.g.~\cite{ellis96}. The location of the divergence, referred to as
the Landau pole, defines the parameter \lmqcd\ which in \oa\ is given
implicitly by $\asq=1/(\beta_0\ln(Q^2/\lmqcdsq))$. The value of
\lmqcd\ is approximately 200~MeV and indicates the scale where
perturbative evolution of \as\ breaks down.

\subsection{ Perturbative QCD Calculations }

The perturbative QCD prediction in \oaa\ for a normalised differential
event shape distribution $dR_{PT}/dy$ of a generic observable $y$
measured at cms energy $\roots=Q$ may be written as
follows~\cite{magnoli90,OPALPR134}:
\begin{equation}
  \frac{dR_{PT}}{dy}= \frac{1}{\sigma_{tot}}\frac{d\sigma}{dy}
               = \frac{dA}{dy}\cf\ash(\mu) + 
  \left(\left(\pi\beta_0\ln(\xmu^2)-\frac{3}{4}\cf\right)2\cf\frac{dA}{dy}
                  +\frac{dB}{dy}\right)\ash^2(\mu)\;\;\;.
\end{equation}
Here $\sigma_{tot}$ is the cross section for the process
$\epem\rightarrow\mathrm{hadrons}$, $dA/dy$ and $dB/dy$ are the \oa\
and the \oaa\ coefficients, respectively, and $\ash=\as/(2\pi)$. The
coefficients $dB/dy$ are in fact a sum of three terms each
proportional to a colour factor:
\begin{equation}
  \frac{dB}{dy}=\cf\left(\frac{dB_{\cf}}{dy}\cf+\frac{dB_{\ca}}{dy}\ca+
                         \frac{dB_{\nf}}{dy}\nf\right)\;\;\;.
\end{equation}
The coefficients $dA/dy$ and $dB_i/dy$, $i=\cf,\ca,\nf$, are obtained
by integrating the \oaa\ QCD matrix elements~\cite{ert} in the \msb\
renormalisation scheme using the program EVENT2~\cite{event2}. A QCD
prediction in \oaa\ is expected to be valid in a region of phase space
where the radiation of a single hard gluon dominates (3-jet region).

For the observables considered here calculations in NLLA have been
performed~\cite{nllathmh,nllabtbw2,nllacp}. The NLLA is valid in a
region of phase space where multiple radiation of soft and collinear
gluons from a system of two hard and back-to-back partons dominates
(2-jet region). In the NLLA the cumulative normalised cross section
$R_{NLLA}(y)$ is written as:
\begin{equation}
  R_{NLLA}(y)= \int_0^y \frac{1}{\sigma_{tot}}\frac{d\sigma}{dy'} dy' =
  C(\as)e^{G(\as,L)}
\end{equation}
with $L=\ln(1/y)$. The functions $C(\as)$ and $G(\as,L)$ are known in
NLLA as:
\begin{equation}
  C(\as) = 1+C_1\ash+C_2\ash^2 \;\;\;\mathrm{and}\;\;\;
  G(\as,L) = \sum_{n=1}^{\infty}\sum_{m=1}^{n+1}G_{nm}\ash^nL^m \simeq
             Lg_1(L\as)+g_2(L\as) \;\;\;.
\end{equation}
The coefficients $C_1$, $G_{11}$, $G_{12}$, $G_{22}$ and $G_{23}$ are
known analytically with complete colour structure while analytical
expressions for the coefficients $C_2$ and $G_{21}$ are not
available. 

The combination of the \oaa\ and the NLLA calculations is a procedure
called matching. Several matching schemes have been used in
measurements of \as, see e.g.~\cite{OPALPR075}. We choose to employ the
so-called \lnr-matching scheme with the following
implementation~\cite{OPALPR075}: 
\begin{eqnarray}
\label{equ_lnr}
\ln R_{PT}(y) & = & Lg_1(L\as) + g_2(L\as) 
         -(G_{11}L+G_{12}L^2)\ash-(G_{22}L^2+G_{23}L^3)\ash^2 \\ \nonumber
         & & +A(y)\ash+(B(y)-\frac{1}{2}A(y)^2)\ash^2\;\;\;.
\end{eqnarray}
The subtraction of the $G_{nm}$ coefficients takes account of the
contributions contained in the NLLA as well as in the \oaa\
calculations. The term $A(y)$ is defined by $A(y)=\int_0^y(dA/dy')dy'$
and equivalently $B(y)=\int_0^y(dB/dy')dy'$. Our choice of the
\lnr-matching scheme is motivated by two arguments, i) it is preferred
theoretically~\cite{nllathmh} as well as
experimentally~\cite{OPALPR075} and ii) it depends only on $G_{nm}$
coefficients known analytically with full colour structure.

\subsection{ Non-perturbative QCD Calculations: Power Corrections }

The model of Dokshitzer, Marchesini and Webber
(DMW)~\cite{dokshitzer95a} treats the effects of gluon radiation at
low energy scales (${\cal O}(\lmqcd)$) where simple perturbative
evolution of \as\ breaks down due to the presence of the Landau
pole. The model assumes that evolution of the strong coupling
\as\ down to energies around and below the Landau pole is
possible. The form of \asmu\ at such low energy scales is a priori
unknown and a non-perturbative parameter is introduced as the 0th
moment over \asmu:
\begin{equation}
\anull= \frac{1}{\mui}\int_0^{\mui}\as(k)dk\;\;\;.
\end{equation}
The quantity \mui\ is referred to as the infrared matching scale
where the non-perturbative and standard perturbative evolution of \as\
are merged, generally taken to be 2~GeV. 

For the observables considered here the power corrections to the
differential distributions have been calculated up to two
loops~\cite{dokshitzer98b,dokshitzer99a}. It turns out that a
distribution $dR/dy$ measured at cms energy $\roots=Q$ can be
described by the shifted perturbative prediction $dR_{PT}/dy$:
\begin{equation}
  \frac{dR}{dy}= \frac{dR_{PT}}{dy}(y-PD_y)\;\;\;.
\label{equ_powc}
\end{equation}
The factor $P$ is universal~\cite{dokshitzer98b} and depends mainly on
the non-perturbative parameter \anull, the infrared matching scale
\mui\ and the cms energy $Q$:
\begin{equation}
  P= \frac{4\cf}{\pi^2}\mil\frac{\mui}{Q}
     \left(\anull(\mui)-\asq-
       2\beta_0\as^2(Q)\left(\ln\frac{Q}{\mui}+\frac{K}{\beta_0}+1
     \right)\right)
\label{equ_p}
\end{equation}
with $K=(67/18-\pi^2/6)\ca-(5/9)\nf$ for the \msb\ renormalisation
scheme. The negative terms containing \asq\ are necessary for a
consistent merging of the strong coupling in the non-perturbative and
the perturbative region.
The quantity \mil\ (Milan factor) stems from two-loop
effects and is given by~\cite{dokshitzer99b}
\begin{equation}
 \mil=1+(1.575\ca-0.104\nf)/(4\pi\beta_0)
\label{equ_milan}
\end{equation}
Its numerical value is 1.49 in standard QCD with $\nf=3$ with an
estimated theoretical uncertainty of 20\%~\cite{dokshitzer98b}.
The factor $D_y$ in equation~(\ref{equ_powc}) is observable
specific and is given by:
\begin{eqnarray}
\label{equ_d}
  D_{\thr} & = & 2 \\ \nonumber
  D_{\cp}  & = & 3\pi \\ \nonumber
  D_{\bt,\bw} & = & \frac{1}{2}\ln\frac{1}{y}+F_y(y,\as(yQ))\;\;\;.
\end{eqnarray}
The functions $F_y$, $y=\bt,\bw$, with known colour structure, describe
additional changes to the distributions referred to as
squeeze~\cite{MovillaFernandez98,dokshitzer99a}.

\section{ Analysis of the Data }
\label{sec_fits}

\subsection{ Event Shape Data }

We use all published data on distributions of our observables $T$,
\cp, \bt\ and \bw\ which include experimental systematic uncertainties
in their errors. The data available in the fits are listed in
table~\ref{tab_data} with references and the ranges which are
considered in the fits. The fit ranges are determined by the following
criteria:
\begin{description}
\item[Data with \bm{\roots\geq\mz}] We choose to
follow the approach of the OPAL
collaboration~\cite{OPALPR075,OPALPR158,OPALPR197,OPALPR303} because
we use the same \oaa+NLLA calculations with the \lnr-matching. The
OPAL collaboration required the experimental and hadronisation
corrections to be reasonably uniform and not strongly model
dependent within the fit ranges. In addition, bins at the edges of fit
ranges with large \chisq\ contributions were removed. The fit
ranges of distributions of the other experiments are adjusted to match
the ones of OPAL as closely as possible. 
\item[Data with \bm{\roots<\mz}]
We use the studies of the recently reanalysed JADE data as a
guideline~\cite{jadenewas,jadec}. However, we require in addition that
i) the fit ranges should extend less far into the 2-jet region than
those used in~\cite{OPALPR075} and ii) the distance between the
extreme 2-jet region and the fit ranges should increase with
decreasing \roots. These requirements reduce the sensitivity of the
analysis to hadronisation corrections which are generally larger in
the 2-jet region than in the 3-jet region. We also demand that no
single bin at the edge of a fit range should have a large \chisq\
contribution.
\end{description}
Within these ranges experimental corrections for limited acceptance
and resolution as well as non-perturbative effects as estimated with
Monte Carlo models are well under control.

\subsection{ Fit Procedure }

The fits are based on equation~(\ref{equ_runas}) for the running of
\as, on equation~(\ref{equ_lnr}) for the perturbative QCD prediction and
on equation~(\ref{equ_powc}) for the power corrections. We vary the
strong coupling \asmz\ with the mass of the \znull\ boson as a
reference scale, one colour factor, i.e. \nf, \ca\ or \cf, and
optionally the non-perturbative parameter \anull\ in the fits. We also
investigate fits where \asmz, the two colour factors \ca\ and \cf\ and
optionally \anull\ are free parameters. Fixed parameters are always
set to the values as expected in standard QCD, i.e. $\nf=5$, $\ca=3$
and $\cf=4/3$. To avoid unphysical results and to stabilise the fits
the free parameters are bounded by $0.01<\asmz<0.3$, $0<\nf<20$,
$0<\ca<10$, $0<\cf<10$ and $0<\anull<10$.

The fit procedure minimises a \chisq\ constructed from the difference
in bin $i$ between the data value $d_i$ and the theoretical prediction
$t_i$ divided by the total error $\sigma_i$:
\begin{equation}
  \chisq= \sum_{i\in\mathrm{fit ranges}} 
              \left(\frac{d_i-t_i}{\sigma_i}\right)^2\;\;\;.
\end{equation}
Possible correlations between bins of a given distribution or between
different distributions are neglected. With the exception of the OPAL
data for \thr, \bt\ and \bw\ at $\roots=\mz$ the data are measured in
bins which are wider than the typical experimental resolution in
order to reduce bin-to-bin migrations. 
Many of the distributions used in this analysis were measured by the
same experiments such that the presence of some correlation between
distributions may be expected. There may also be correlations between
the data points of any given distribution due to common systematic
uncertainties. However, information about such correlations is not
available in the references for the data (see table~\ref{tab_data}). 

The errors on the fitted parameters are calculated in the fit
procedure from the diagonal elements of the covariance matrix after
the fit has converged (``fit error''). These errors contain the
contributions from statistical fluctuations and systematic
uncertainties quoted by the individual experiments.

We follow two alternatives to study the influence of non-perturbative
effects on our fits. Firstly, we fix the non-perturbative parameter
\anull\ at a previously measured value and repeat the fits with
\anull\ varied by its total errors. Secondly, we allow \anull\ to vary
in the fit. The fixed value of \anull\ is taken from~\cite{jadec},
$\anull=0.473^{+0.058}_{-0.041}$. However, some modifications had to
be applied. The measurement of~\cite{jadec} used an erroneous value of
the Milan factor, $\mil=1.794$ instead of 1.49, both for $\nf=3$. From
equation~(\ref{equ_p}) we infer that it is sufficient to scale \anull\
by the ratio of the two values for \mil\ as a correction. In our study
we vary the colour factor \nf\ and it would therefore be inconsistent
to use two different values of \nf, i.e. $\nf=5$ in the perturbative
part and $\nf=3$ in the non-perturbative part of the QCD prediction
for massless quarks. To find our standard value of \anull\ we scale it
by the ratio of values for the Milan factor determined with $\nf=3$
and $\nf=5$, respectively. Our final value is $\anull=0.543\pm0.058$
with symmetrised errors.

\subsection{ Effects of \bm{ \bbbar } events at low \bm{ \roots } }
\label{sec_bmass}

The presence of events from the reaction $\epem\rightarrow\bbbar$ at
low cms energies \roots\ can distort the event shape distributions,
because the effects of weak decays of heavy B-hadrons on the topology
of hadronic events cannot anymore be neglected. An additional
potential problem arises from comparing QCD calculations based on
massless quarks with data containing massive quarks at \roots\ close
to the production threshold.

At $\roots\ll\mz$ \bbbar\ events constitute about 9\% of the total
event samples. Ideally one would correct the data experimentally by
identifying \bbbar\ events and removing them from the sample. However,
since we have only published event shape data without information on
specific quark flavours we resort to a correction based on Monte
Carlo simulations. We generate samples of $10^6$ events at each
\roots\ with the JETSET~7.4 program~\cite{jetset3} with the parameter
set given in~\cite{OPALPR141}. For each event shape observable we
build the ratio of distributions calculated with u, d, s and c quark
events to those calculated with all events. This ratio is multiplied
with the bin contents of the data to obtain corrected
distributions. This procedure is applied to all data at
$\roots<\mz$. It was verified that the simulation provides an adequate
description of the data at all values of $\roots<\mz$.

Systematic effects due to uncertainties in the Monte Carlo parameters
are expected to be small for the ratio except for those parameters
which only affect the \bbbar\ events in the samples. The most
important such parameter is the value of \epsb\ in the Peterson
fragmentation function~\cite{peterson83} which controls the
fragmentation of b quarks in the simulation. Threshold effects at low
\roots\ which depend on the value of the b-quark mass in the
simulation are found to be negligible for the fit results.

\subsection{ Systematic Variations }

As systematic variations we consider the following changes in the
analysis:
\begin{description}
\item[Renormalisation scale] The standard fits are carried out with
the renormalisation scale parameter $\xmu=1$. The dependence on the
renormalisation scale is investigated by repeating the fits with
$\xmu=0.5$ and $\xmu=2.0$~\cite{OPALPR075}. Deviations of the fit
results w.r.t. the standard fits are taken as asymmetric
uncertainties.
\item[Power corrections (\bm{ \anull } fixed)] The standard fits with
fixed \anull\ employ $\anull=0.543$, as explained above. The fits are
repeated with \anull\ varied by its error, i.e. $\anull=0.485$ and
$\anull=0.601$.  Again, deviations of the fit results w.r.t. the
standard fits are taken as asymmetric uncertainties.
\item[Power corrections (\bm{ \anull } free)] The standard fits use
the Milan factor as given by equation~(\ref{equ_milan}). The fits are
repeated with the Milan factor scaled by a factor of 0.8 and 1.2,
respectively, and deviations w.r.t. results of the standard fits are
considered as asymmetric uncertainties. We also change the value of
the infrared matching scale \mui\ from its standard value of 2~GeV
to 1~GeV and 3~GeV, respectively,  and repeat the fits. Deviations
w.r.t. the standard results are counted as asymmetric uncertainties,
except for \anull\ where changing \mui\ corresponds to a redefinition
of the non-perturbative parameter \anull.  
\item[Fragmentation of b quarks] The standard analysis is carried out
with corrected data at $\roots<\mz$ based on the JETSET tuning
of~\cite{OPALPR141} as explained in section~\ref{sec_bmass}. The value
of the JETSET parameter \epsb\ is varied around its central value
$\epsb=0.0038\pm0.0010$ by adding or subtracting its error and the
analysis including correction of the data at $\roots<\mz$ is
repeated. Deviations w.r.t. the standard results are considered as
asymmetric uncertainties.
\item[Experimental uncertainties] We change the composition of the
data sets for the fits in two ways, i) the data measured at
$\roots=\mz$ are removed and ii) only data measured at $\roots\geq\mz$
are used. The larger of the two deviations w.r.t. the standard result
observed with the data set i) or ii) is considered as the systematic
uncertainty from variations in the input data.
\end{description}
The systematic uncertainties from the renormalisation scale, the power
corrections with \anull\ either fixed or free, the fragmentation of b
quarks and from the input data are added in quadrature with the error
from the fit to arrive at the total error.

\subsection{ Fit Results }

\begin{figure}[!htb]
\begin{center}
\includegraphics[width=\textwidth]{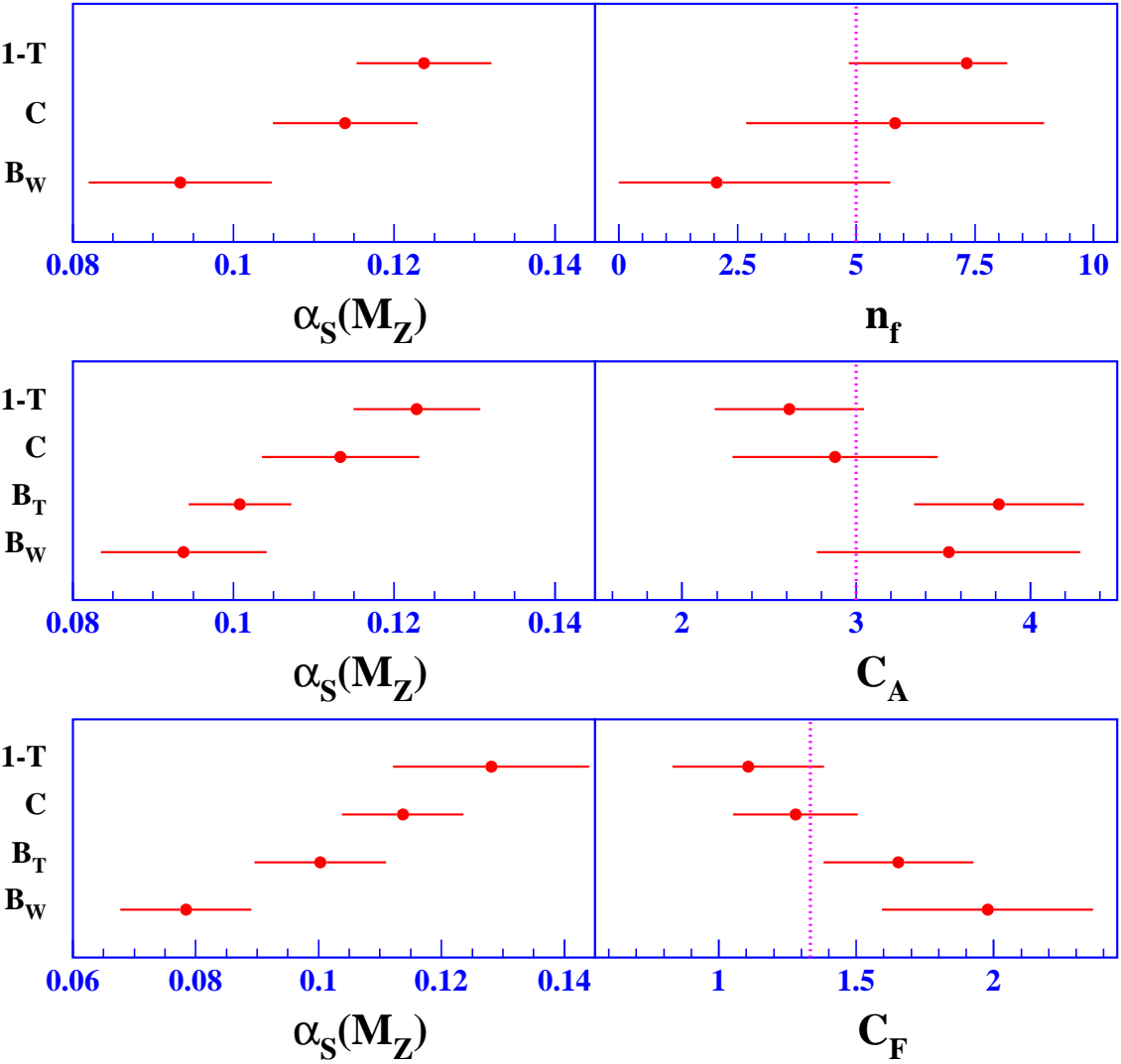}
\caption[ bla ]{ The figures present results of fits to \asmz\ and
one of the colour factors \nf, \ca\ or \cf\ with observables as
given on the vertical axis. The vertical dotted lines
indicate the expectation from standard QCD for the colour factor. }
\label{fig_plot}
\end{center}
\end{figure}

\begin{figure}[!htb]
\begin{center}
\includegraphics[width=\textwidth]{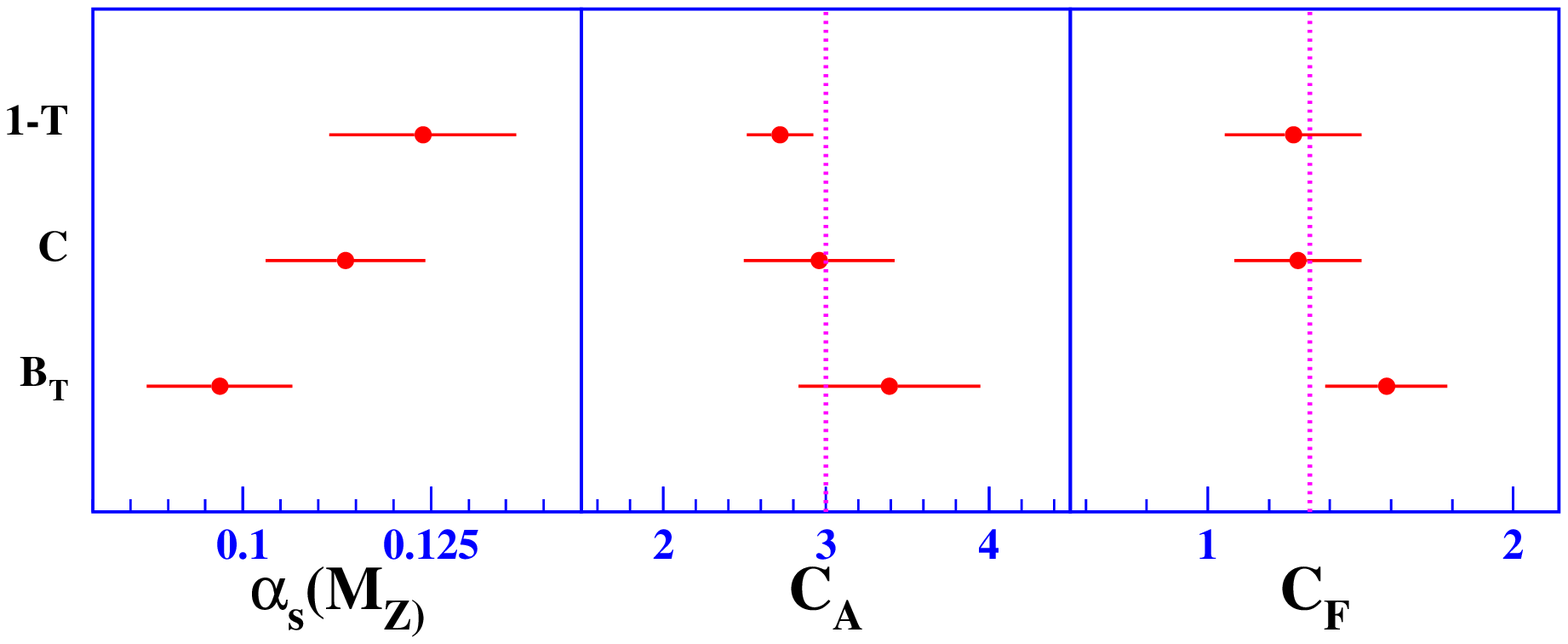}
\caption[ bla ]{ The figure presents results from fits to \asmz\ and
the colour factors \ca\ and \cf\ with observables as given on the
vertical axis. The error bars show total uncertainties. The vertical
dotted lines indicate the expectations from standard QCD for the
colour factors. }
\label{fig_plotcacf}
\end{center}
\end{figure}

\begin{figure}[!htb]
\begin{center}
\includegraphics[width=\textwidth]{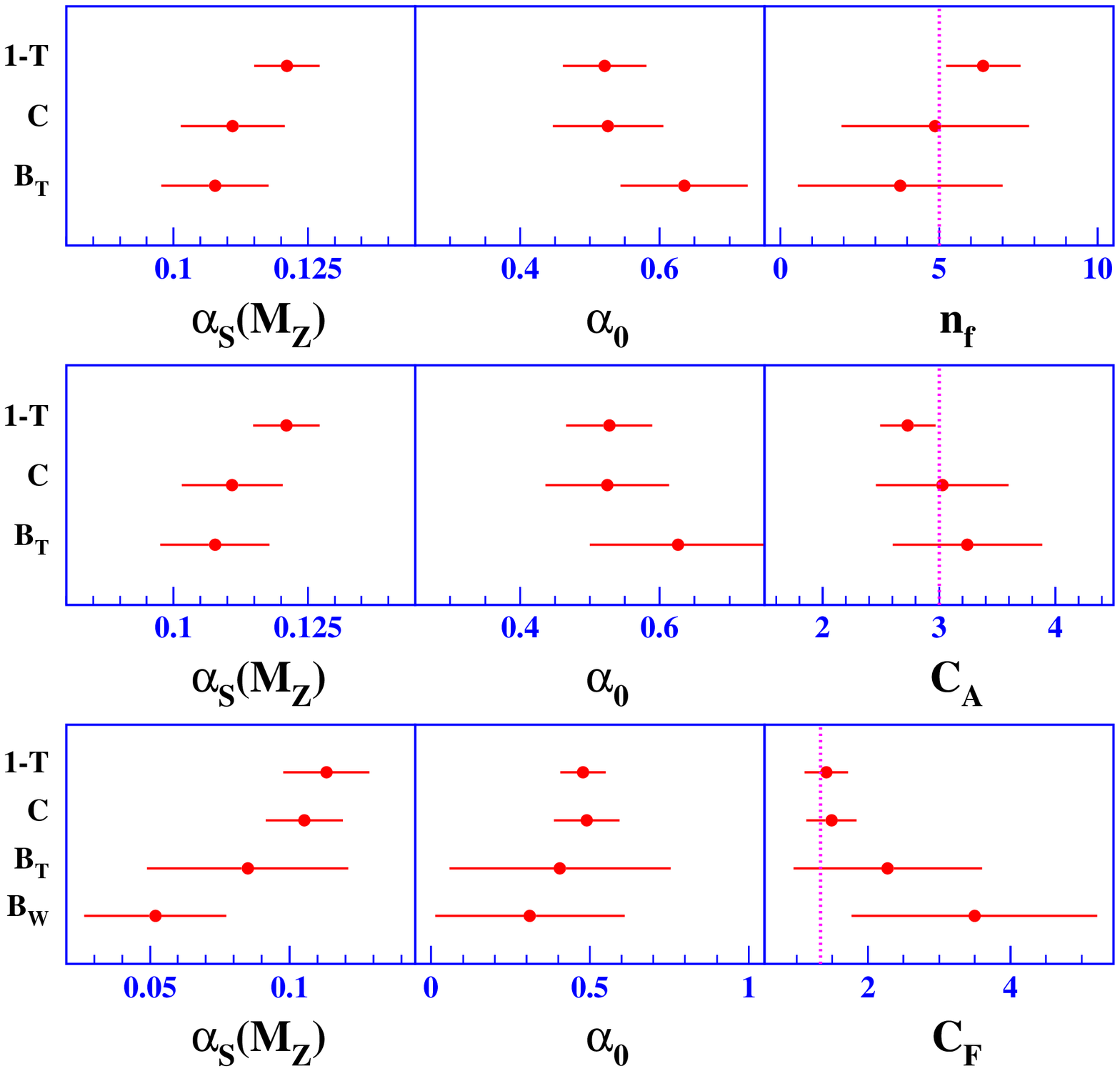}
\caption[ bla ]{ The figures present results from fits to \asmz,
\anull\  and one of the colour factors \nf, \ca\ or \cf\ with
observables as given on the vertical axis. The error bars show total
uncertainties. The vertical dotted lines indicate the expectation from
standard QCD for the colour factor. }
\label{fig_plota0}
\end{center}
\end{figure}

The tables~\ref{tab_res_nf}, \ref{tab_res_ca} and \ref{tab_res_cf}
present the results of two-parameter fits to \asmz\ and \nf, \ca\ or
\cf, respectively, with \anull\ fixed. The results of three-parameter
fits to \asmz, \ca\ and \cf\ are given in
table~\ref{tab_res_cacf}. The results of three-parameter fits to \asmz,
\anull\ and one of the colour factors \nf, \ca\ or \cf\ are shown in
tables~\ref{tab_res_nf_a0free}, \ref{tab_res_ca_a0free} and
\ref{tab_res_cf_a0free}, respectively, while
table~\ref{tab_res_cacf_a0free} displays the results of four-parameter
fits to \asmz, \anull, \ca\ and \cf. The rows labelled ``hadr.''
contain the quadratic sum of the uncertainties from the power
corrections and the fragmentation of b-quarks. The power correction
uncertainties are the dominating contribution in all cases. The tables
show symmetrised errors except for the fit to \asmz\ and \nf\ where
the variation of \anull\ leads to a large difference between the
positive and negative error. Figures~\ref{fig_plot},
\ref{fig_plotcacf} and \ref{fig_plota0} give a graphic display of the
results.

We observe stable fits in all cases for the observables \thr\ and
\cp. The fit results from these observables for the colour factors are
generally in good agreement with the expectations from standard QCD
within their total uncertainties. The results for \asmz\ and \anull\
are also generally consistent with previous measurements within their
total uncertainties~\cite{bethke00a,jadec}. The fit of \nf\ and \asmz\
with \thr\ is consistent with an earlier analysis using similar
data~\cite{hebbeker93}. The fit of \asmz, \anull, \ca\ and \cf\ with
\cp\ converges but has such large fit errors that sensitivity to the
colour factors is essentially lost; consequently we do not show the
results. 

With the other two observables, \bt\ and \bw, we also find generally
consistent results with standard QCD and previous
measurements. However, fits of \bt\ to \asmz\ and \nf\ and fits of of 
\bw\ to \asmz, \ca\ and \cf\ as well as to \asmz\ and single colour
factors with \anull\ free are unstable and we do not show these results.
Our interpretation is that the QCD calculations with power corrections
for \thr\ and \cp\ provide an adequate description of the data while
there are still effects in the data for \bt\ and \bw\ which are not
well described by the \oaa+NLLA QCD calculations with power
corrections~\cite{OPALPR075,jadenewas,jadec,movilla00a}.

The values of \chisqd\ are smaller than one in all cases, except the
fit of \asmz\ and \ca\ with \bt\ where $\chisqd=1.01$. The small
values of \chisqd\ are consistent with the good description of the
data by the fitted predictions and probably indicate the presence of
correlations between the data points used in a given fit. 

We find that in all fits with \anull\ free the variation of the
Milan factor leads to significantly smaller systematic uncertainties
than the variation of \anull\ in fits with \anull\ fixed.  The fitted
values of \anull\ vary over ranges compatible with the total error of
the fixed value of \anull\ when the Milan factor is changed. This is
consistent with~\cite{jadec} where the error on \anull\ was dominated
by the variation of the Milan factor. We conclude that the ratio of
non-perturbative and perturbative contributions is well constrained by
the data such that changes in the Milan factor are compensated by
similar changes with the opposite sign in \anull. As a consequence we
find that the hadronisation systematic uncertainties of fits to the
colour factors are reduced when \anull\ is a free parameter while the
fit errors increase due to the presence of an extra free parameter in
the fit. 

We observe that the experimental uncertainties estimated by repeating
the fits with reduced data sets are larger in the fits with \anull\
free. This effect is likely to be a reflection of the increased fit
errors. We choose to keep the larger uncertainties since we cannot
rule out the possibility that there are systematic effects in the data
to which the fits with \anull\ free are more sensitive than the fits
with \anull\ fixed.

In simultaneous fits with several free parameters correlations between
them may influence the results. In table~\ref{tab_corr} we show the
correlations observed in the standard fits with the observable
\thr. In addition in the first row of table~\ref{tab_corr} the
correlation observed in a fit with only \asmz\ and \anull\ as free
parameters is displayed. The presence of large correlations in some
fits causes increased fit errors and may also contribute to the
experimental uncertainties. We find a similar pattern of correlations
with the other observables \cp, \bt\ and \bw.

\subsection{ Combination of Results }

We construct combined results only from fits to \thr\ and \cp, because
we found these to be stable in all cases. We choose to base the
combination for individual colour factors on the fits with \anull\
free since these results should be less biased by input values
measured assuming standard QCD. They also have smaller hadronisation
uncertainties as discussed above.  We build unweighted averages of the
fit results for the standard analysis and for all systematic
variations. As the fit error we choose the smaller of the two
individual fit errors and we then construct the total error in the
same way as for an individual observable. In this way correlations
between systematic variations of the observables are taken into
account. We find as our final results for the individually measured
colour factors:
\begin{eqnarray} 
\label{equ_indres}
\nf & = & 5.64\pm0.79\fit\pm0.32\scale\pm0.48\had\pm0.93\exptl \\ \nonumber
    & = & 5.64\pm1.35, \\ \nonumber
\ca & = & 2.88\pm0.16\fit\pm0.06\scale\pm0.10\had\pm0.18\exptl \\ \nonumber
    & = & 2.88\pm0.27\;\;\;\mathrm{and} \\ \nonumber
\cf & = & 1.45\pm0.21\fit\pm0.04\scale\pm0.04\had\pm0.16\exptl \\ \nonumber
    & = & 1.45\pm0.27. \\ \nonumber
\end{eqnarray}
The total uncertainties for all three colour factors are
dominated by the fit error and the experimental uncertainties. The
results are in good agreement with the expectation from standard QCD
with five active quark flavours. The results for \asmz\ and \anull\
from the combination are consistent with the individual fits and with
previous measurements~\cite{bethke00a,jadec}, see
table~\ref{tab_combres}. 

In the case of \nf\ the combined result has a slightly larger total
uncertainty than the individual result based on the observable
\thr\ shown in table~\ref{tab_res_nf_a0free}. This is mainly due to
the increased experimental uncertainty of the combined result which in
turn is dominated by the experimental uncertainty of the fit with
\cp. For the other colour factors \ca\ and \cf\ the total
uncertainties of the combined results are somewhat smaller than the
total uncertainties of the individual results. These observations are
consistent, since we do not expect the total uncertainties to be
greatly reduced by a combination of individual results due to the
large correlations between the event shape distributions.

A combined result for the results of the simultaneous fits of \ca\ and
\cf\ may only be constructed from the fits with \anull\ fixed, because
in the fits with \anull\ free we obtained usable fits only for the
observable \thr. Applying the same procedure as outlined above yields
the following results for combining the results of the fits with the
observables \thr\ and \cp\ shown in table~\ref{tab_res_cacf}:
\begin{eqnarray}
\label{equ_simres}
\ca & = & 2.84\pm0.13\fit\pm0.06\scale\pm0.11\had\pm0.15\exptl \\ \nonumber
    & = & 2.84\pm0.24\;\;\;\mathrm{and} \\ \nonumber
\cf & = & 1.29\pm0.07\fit\pm0.17\had\pm0.02\exptl \\ \nonumber
    & = & 1.29\pm0.18. \\ \nonumber
\end{eqnarray}
The scale uncertainty of the combined result for \cf\ is smaller than
0.01 consistent with the small scale uncertainties of the individual
results, see table~\ref{tab_res_cacf}. The results for \ca\ and \cf\
are in good agreement with the combined results discussed above, see
equation~(\ref{equ_indres}). The combined result for \asmz\ is in
agreement with the individual results and also with~\cite{bethke00a},
see table~\ref{tab_combres}.

\begin{figure}[!htb]
\begin{center}
\includegraphics[width=\textwidth]{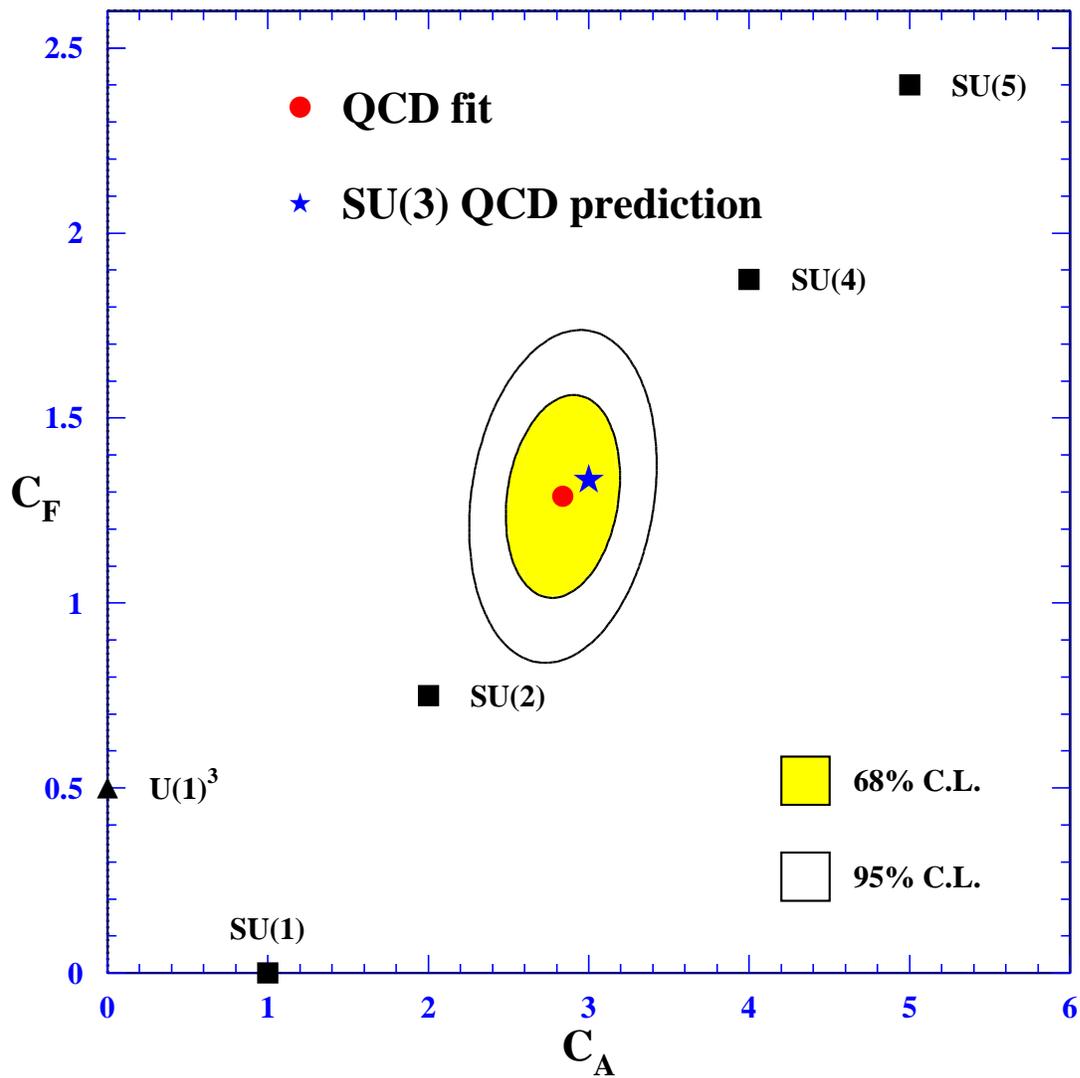}
\caption[ bla ]{ The figure presents the combined results for the
colour factors \ca\ and \cf\ from fits to \asmz, \ca\ and \cf\ based
on the observables \thr\ and \cp. The square and triangle symbols
indicate the expectations for \ca\ and \cf\ for different symmetry
groups. }
\label{fig_plotcacfcont}
\end{center}
\end{figure}

The averaged correlation coefficient from the fits is
$\rho_{\ca-\cf}=-0.89$. However, we observe that variations of the
results for \ca\ and \cf\ are positively correlated when \anull\ is
changed. In order to find a conservative estimate of the correlation
coefficient we construct a covariance matrix by summing i) the
covariance matrix from the fit and ii) one covariance matrix for each
systematic uncertainty. The covariance matrices ii) are constructed
from the systematic uncertainties, symmetrised if necessary, and
$\rho_{\ca-\cf}=-0.89$, except for the systematic uncertainty from the
variation of \anull\ where $\rho_{\ca-\cf}=1.0$ is used. The resulting
total correlation coefficient is
$\rho_{\ca-\cf}=0.19$. Figure~\ref{fig_plotcacfcont} presents the
combined results which are in good agreement with standard QCD within
the uncertainties. Some other possible gauge groups are indicated
including U(1)$^3$, an abelian QCD with three quark colours and colour
neutral gluons. Our results exclude all shown alternatives to SU(3) as
the gauge group of QCD at more than 95\% confidence level.

\section{ Summary and Conclusions }
\label{sec_summ}

We have studied fits of \oaa+NLLA QCD predictions for distributions of
the event shape observables \thr, \cp, \bt\ and \bw\ with power
correction calculations to model hadronisation effects to data
measured at cms energies ranging from 14~GeV to 189~GeV. In these fits
we varied simultaneously the strong coupling \asmz, one of the QCD
colour factors \ca\ or \cf, or the the number active quarks \nf, and
in some cases also \anull, the free parameter of the power correction
calculations. We investigated in addition fits, where \asmz, \ca\ and
\cf\ and optionally \anull\ were varied.

We found stable fits in all cases with the observables \thr\ and \cp\
while some fits with \bt\ and in particular \bw\ appear unreliable. We
take this as an indication that the current \oaa+NLLA QCD calculations
with power corrections for \bt\ and \bw\ describe the data not as well
as the same calculations for \thr\ and \cp. 

We observed that the variation of the Milan factor \mil\ and the
infrared matching scale \mui\ in the fits with \anull\ free gave rise
to smaller systematic uncertainties than the systematic variation of
\anull\ in the fits with \anull\ fixed. Our conclusion is that the
relative contributions of the perturbative \oaa+NLLA QCD calculations
and the power correction terms to the total predictions are quite well
constrained by the data. 

A combination of the individual results of the fits with \asmz, an
individual colour factor and \anull\ as free parameters 
with the observables \thr\ and \cp\ yields our final results:
\begin{eqnarray} \nonumber
\nf & = & 5.64\pm1.35, \\ \nonumber
\ca & = & 2.88\pm0.27\;\;\;\mathrm{and} \\ \nonumber
\cf & = & 1.45\pm0.27. \\ \nonumber
\end{eqnarray}
The combination of results from simultaneous fits of \ca, \cf\ and
\asmz\ for \anull\ fixed with the observables \thr\ and \cp\ gives
\begin{equation}
\ca = 2.84\pm0.24\;\;\;\mathrm{and}\;\;\;
\cf = 1.29\pm0.18
\end{equation}
with total correlation coefficient $\rho_{\ca-\cf}=0.19$.  These
results are in good agreement with the expectation from standard QCD
based on the SU(3) symmetry group for \epem\ annihilation data at high
energies, i.e. $\nf=5$ with $\tf=1/2$, $\ca=3$ and $\cf=4/3$. There is
also good agreement between the individual and the simultaneous
measurements of \ca\ and \cf. All combined results for \asmz\ and
\anull\ are in agreement with previous measurements. We found that the
total uncertainties of the combined results are approximately of the
same size as the total uncertainties of the individual results. This
is consistent, because the event shape distributions are correlated
with each other and the total uncertainties are dominated by the fit
errors and experimental uncertainties.

We present our final conclusions from two points of view. Firstly, we
assume that the power correction calculations are a good model of
hadronisation effects in event shape distributions. In this case we
have performed a complementary test of the gauge structure of QCD with
competitive uncertainties on the measurements of the QCD colour
factors compared to other
analyses~\cite{dienes00,aleph4jet2,4jetdelphi3,OPALPR113}. Secondly,
under the assumption that QCD with SU(3) group symmetry is the
correct theory of strong interactions, our analysis provides a
successful consistency check of the power correction calculations.

\clearpage

\section*{ Tables }

\renewcommand{\exptl}   {exp.}

\begin{table}[!htb]
\begin{center}
\begin{tabular}{|r|l||c|c|c|c|}
\hline
\roots & Experiment & \thr & \cp & \bt & \bw \\ \hline\hline
189    & L3 \cite{l3lep2data1} & 
0.05-0.30 &  0.15-0.60 & 0.08-0.26 & 0.045-0.195 \\
       & OPAL \cite{OPALPR303} & 
0.04-0.30 & 0.18-0.60 & 0.075-0.25 & 0.05-0.20 \\ \hline
183    & DELPHI \cite{delphilep2data1} & 
0.04-0.28 & 0.16-0.64 & 0.07-0.24 & 0.05-0.20 \\
       & L3 \cite{l3lep2data1} & 
0.05-0.30 & 0.15-0.60 & 0.08-0.26 & 0.045-0.195 \\
       & OPAL \cite{OPALPR303} & 
0.04-0.30 & 0.18-0.60 & 0.075-0.25 & 0.05-0.20 \\ \hline
172    & DELPHI \cite{delphilep2data1} & 
0.04-0.24 & 0.16-0.64 & 0.08-0.27 & 0.04-0.17 \\
       & L3 \cite{l3lep2data1} & 
0.05-0.30 & 0.15-0.60 & 0.08-0.26 & 0.045-0.195 \\
       & OPAL \cite{OPALPR303} & 
0.04-0.30 & 0.18-0.60 & 0.075-0.25 & 0.05-0.20 \\ \hline
161    & DELPHI \cite{delphilep2data1} & 
0.04-0.24 & 0.16-0.64 & 0.08-0.27 & 0.04-0.17 \\
       & L3 \cite{l3lep2data1} & 
0.05-0.30 & 0.15-0.60 & 0.08-0.26 & 0.045-0.195 \\
       & OPAL \cite{OPALPR197} & 
0.04-0.30 & 0.18-0.60 & 0.075-0.25 & 0.05-0.20 \\ \hline
133    & ALEPH \cite{alephas133} & 0.04-0.30 & & & \\
       & DELPHI \cite{delphilep2data1} & 
0.04-0.24 & 0.16-0.64 & 0.08-0.27 & 0.04-0.17 \\
       & L3 \cite{l3lep2data1} & 
0.05-0.25 & 0.15-0.64 & 0.08-0.26 & 0.045-0.195 \\
       & OPAL \cite{OPALPR158} & 
0.04-0.30 & 0.18-0.60 & 0.075-0.25 & 0.05-0.20 \\ \hline
91     & ALEPH \cite{alephlep1data} & 
0.06-0.30 & 0.20-0.64 & & \\
       & DELPHI \cite{delphilep1data} & 
0.06-0.30 & 0.20-0.64 & 0.09-0.24 & 0.07-0.17 \\
       & L3 \cite{l3lep1data} & 
0.065-0.29 & 0.22-0.64 & & \\
       & OPAL \cite{OPALPR075,OPALPR054} & 
0.06-0.30 & 0.20-0.64 & 0.09-0.23 & 0.07-0.17 \\
       & SLD \cite{sldnlla} & 
0.06-0.26 & 0.24-0.64 & 0.08-0.26 & 0.08-0.20 \\ \hline
55     & AMY \cite{amydata} & 0.10-0.30 & & & \\ \hline
44     & JADE \cite{jadenewas,jadec} & 
0.08-0.30 & 0.24-0.58 & 0.10-0.24 & 0.08-0.18 \\
       & TASSO \cite{tassodata} & 0.08-0.28 & & & \\ \hline
35     & JADE \cite{jadenewas,jadec} & 
0.08-0.30 & 0.24-0.58 & 0.10-0.24 & 0.08-0.18 \\
       & TASSO \cite{tassodata} & 0.08-0.28 & & & \\ \hline
29     & HRS \cite{hrsdata} & 0.10-0.30 & & & \\ 
       & MARKII \cite{markiidata} & 0.10-0.30 & & & \\ \hline
22     & TASSO \cite{tassodata} & 0.10-0.28 & & & \\ \hline
14     & TASSO \cite{tassodata} & 0.12-0.28 & & & \\
\hline
\end{tabular}
\caption[ bla ]{ The sources of the data and the fit ranges 
for the observables \thr, \cp, \bt\ and \bw\ are shown. The cms energy
\roots\ at which the experiments analysed their data is given in GeV. }
\label{tab_data}
\end{center}
\end{table}

\begin{table}[!htb]
\begin{center}
\begin{tabular}{|r||c|c||c|c||c|c||c|c||} \hline
 & \multicolumn{2}{c||}{\thr} & \multicolumn{2}{c||}{\cp} & \multicolumn{2}{c||}{\bw} \\ \cline{1-7}\cline{1-7}
 & \nf & \asmz & \nf & \asmz & \nf & \asmz \\ \cline{1-7}
 \stan &  7.32 & 0.124 &  5.82 & 0.114 &  2.06 & 0.093 \\ \cline{1-7}
 \fite & $\pm 0.49$ & $\pm0.002$ & $\pm 0.52$ & $\pm0.002$ & $\pm 1.04$ & $\pm0.003$ \\ \cline{1-7}
 \chisqd & \multicolumn{2}{c||}{ 130.8/236 } & \multicolumn{2}{c||}{ 104.0/155 } & \multicolumn{2}{c||}{ 109.0/121 } \\ \cline{1-7}
 \rensc & $\pm 0.11$ & $\pm0.003$ & $\pm 0.55$ & $\pm0.002$ & $\pm 1.33$ & $\pm0.002$ \\ \cline{1-7}
 \hadr & \rule{0pt}{2.2ex} $^{+0.56}_{-2.39}$ & $\pm0.007$ & $\pm 2.30$ & $\pm0.006$ & $\pm 1.47$ & $\pm0.003$ \\ \cline{1-7}
 \exptl & $\pm 0.42$ & $\pm0.003$ & $\pm 1.95$ & $\pm0.006$ & $\pm 2.89$ & $\pm0.010$ \\ \cline{1-7}
 \tote & \rule{0pt}{2.2ex} $^{+0.86}_{-2.48}$ & $\pm0.008$ & $\pm 3.14$ & $\pm0.009$ & $\pm 3.66$ & $\pm0.011$ \\ \cline{1-7}
\end{tabular}
\caption[ bla ]{
Results are shown for fits with the observables \thr, \cp\ and \bw\ to \asmz\ and \nf. }
\label{tab_res_nf}
\end{center}
\end{table}

\begin{table}[!htb]
\begin{center}
\begin{tabular}{|r||c|c||c|c||c|c||c|c||} \hline
 & \multicolumn{2}{c||}{\thr} & \multicolumn{2}{c||}{\cp} & \multicolumn{2}{c||}{\bt} & \multicolumn{2}{c||}{\bw} \\ \cline{1-9}\cline{1-9}
 & \ca & \asmz & \ca & \asmz & \ca & \asmz & \ca & \asmz \\ \cline{1-9}
 \stan &  2.62 & 0.123 &  2.88 & 0.113 &  3.82 & 0.101 &  3.53 & 0.094 \\ \cline{1-9}
 \fite & $\pm 0.08$ & $\pm0.002$ & $\pm 0.07$ & $\pm0.002$ & $\pm 0.10$ & $\pm0.002$ & $\pm 0.16$ & $\pm0.003$ \\ \cline{1-9}
 \chisqd & \multicolumn{2}{c||}{ 128.8/236 } & \multicolumn{2}{c||}{ 103.7/155 } & \multicolumn{2}{c||}{ 139.6/138 } & \multicolumn{2}{c||}{ 105.9/121 } \\ \cline{1-9}
 \rensc & $\pm 0.03$ & $\pm0.004$ & $\pm 0.08$ & $\pm0.003$ & $\pm 0.21$ & $\pm0.003$ & $\pm 0.21$ & $\pm0.001$ \\ \cline{1-9}
 \hadr & $\pm 0.42$ & $\pm0.006$ & $\pm 0.49$ & $\pm0.008$ & $\pm 0.36$ & $\pm0.004$ & $\pm 0.23$ & $\pm0.003$ \\ \cline{1-9}
 \exptl & $\pm 0.07$ & $\pm0.003$ & $\pm 0.31$ & $\pm0.005$ & $\pm 0.23$ & $\pm0.003$ & $\pm 0.67$ & $\pm0.009$ \\ \cline{1-9}
 \tote & $\pm 0.43$ & $\pm0.008$ & $\pm 0.59$ & $\pm0.010$ & $\pm 0.49$ & $\pm0.006$ & $\pm 0.76$ & $\pm0.010$ \\ \cline{1-9}
\end{tabular}
\caption[ bla ]{
Results are shown for fits with the observables \thr, \cp, \bt\ and \bw\ to \asmz\ and \ca. }
\label{tab_res_ca}
\end{center}
\end{table}

\begin{table}[!htb]
\begin{center}
\begin{tabular}{|r||c|c||c|c||c|c||c|c||} \hline
 & \multicolumn{2}{c||}{\thr} & \multicolumn{2}{c||}{\cp} & \multicolumn{2}{c||}{\bt} & \multicolumn{2}{c||}{\bw} \\ \cline{1-9}\cline{1-9}
 & \cf & \asmz & \cf & \asmz & \cf & \asmz & \cf & \asmz \\ \cline{1-9}
 \stan &  1.11 & 0.128 &  1.28 & 0.114 &  1.65 & 0.100 &  1.98 & 0.078 \\ \cline{1-9}
 \fite & $\pm 0.06$ & $\pm0.004$ & $\pm 0.03$ & $\pm0.002$ & $\pm 0.03$ & $\pm0.002$ & $\pm 0.09$ & $\pm0.003$ \\ \cline{1-9}
 \chisqd & \multicolumn{2}{c||}{ 133.4/236 } & \multicolumn{2}{c||}{ 103.6/155 } & \multicolumn{2}{c||}{ 131.3/138 } & \multicolumn{2}{c||}{  87.8/121 } \\ \cline{1-9}
 \rensc & $\pm 0.05$ & $\pm0.004$ & $\pm 0.05$ & $\pm0.003$ & $\pm 0.01$ & $\pm0.003$ & $\pm 0.07$ & $\pm0.001$ \\ \cline{1-9}
 \hadr & $\pm 0.24$ & $\pm0.013$ & $\pm 0.22$ & $\pm0.009$ & $\pm 0.21$ & $\pm0.006$ & $\pm 0.24$ & $\pm0.006$ \\ \cline{1-9}
 \exptl & $\pm 0.10$ & $\pm0.008$ & $\pm 0.04$ & $\pm0.001$ & $\pm 0.17$ & $\pm0.008$ & $\pm 0.27$ & $\pm0.008$ \\ \cline{1-9}
 \tote & $\pm 0.28$ & $\pm0.016$ & $\pm 0.23$ & $\pm0.010$ & $\pm 0.27$ & $\pm0.011$ & $\pm 0.38$ & $\pm0.011$ \\ \cline{1-9}
\end{tabular}
\caption[ bla ]{
Results are shown for fits with the observables \thr, \cp, \bt\ and \bw\ to \asmz\ and \cf. }
\label{tab_res_cf}
\end{center}
\end{table}

\begin{table}[!htb]
\begin{center}
\begin{tabular}{|r||c|c|c||c|c|c||} \hline
 & \multicolumn{3}{c||}{\thr} & \multicolumn{3}{c||}{\cp} \\ \cline{1-7}\cline{1-7}
 & \ca & \cf & \asmz & \ca & \cf & \asmz \\ \cline{1-7}
 \stan &  2.72 &  1.28 & 0.124 &  2.96 &  1.30 & 0.114 \\ \cline{1-7}
 \fite & $\pm 0.13$ & $\pm 0.07$ & $\pm0.002$ & $\pm 0.18$ & $\pm 0.08$ & $\pm0.002$ \\ \cline{1-7}
 \chisqd & \multicolumn{3}{c||}{ 127.9/235 } & \multicolumn{3}{c||}{ 103.6/154 } \\ \cline{1-7}
 \rensc & $\pm 0.04$ & $\pm 0.00$ & $\pm0.004$ & $\pm 0.08$ & $\pm 0.00$ & $\pm0.003$ \\ \cline{1-7}
 \hadr & $\pm 0.10$ & $\pm 0.17$ & $\pm0.010$ & $\pm 0.12$ & $\pm 0.16$ & $\pm0.009$ \\ \cline{1-7}
 \exptl & $\pm 0.11$ & $\pm 0.13$ & $\pm0.005$ & $\pm 0.40$ & $\pm 0.10$ & $\pm0.004$ \\ \cline{1-7}
 \tote & $\pm 0.20$ & $\pm 0.22$ & $\pm0.012$ & $\pm 0.46$ & $\pm 0.21$ & $\pm0.011$ \\ \cline{1-7}
 & \multicolumn{3}{c||}{\bt} \\ \cline{1-4}\cline{1-4}
 & \ca & \cf & \asmz \\ \cline{1-4}
 \stan &  3.39 &  1.58 & 0.097 \\ \cline{1-4}
 \fite & $\pm 0.21$ & $\pm 0.06$ & $\pm0.002$ \\ \cline{1-4}
 \chisqd & \multicolumn{3}{c||}{ 128.0/137 } \\ \cline{1-4}
 \rensc & $\pm 0.10$ & $\pm 0.01$ & $\pm0.002$ \\ \cline{1-4}
 \hadr & $\pm 0.16$ & $\pm 0.19$ & $\pm0.008$ \\ \cline{1-4}
 \exptl & $\pm 0.48$ & $\pm 0.02$ & $\pm0.005$ \\ \cline{1-4}
 \tote & $\pm 0.56$ & $\pm 0.20$ & $\pm0.010$ \\ \cline{1-4}
\end{tabular}
\caption[ bla ]{
Results are shown for fits with the observables \thr, \cp\ and \bt\ to
\asmz, \ca\ and \cf. }
\label{tab_res_cacf}
\end{center}
\end{table}

\begin{table}[!htb]
\begin{center}
\begin{tabular}{|r||c|c|c||c|c|c||} \hline
 & \multicolumn{3}{c||}{\thr} & \multicolumn{3}{c||}{\cp} \\ \cline{1-7}\cline{1-7}
 & \nf & \asmz & \anull & \nf & \asmz & \anull \\ \cline{1-7}
 \stan &  6.39 & 0.121 & 0.521 &  4.88 & 0.111 & 0.526 \\ \cline{1-7}
 \fite & $\pm 0.79$ & $\pm0.003$ & $\pm0.015$ & $\pm 1.34$ & $\pm0.004$ & $\pm0.023$ \\ \cline{1-7}
 \chisqd & \multicolumn{3}{c||}{ 127.9/235 } & \multicolumn{3}{c||}{ 103.4/154 } \\ \cline{1-7}
 \rensc & $\pm 0.21$ & $\pm0.004$ & $\pm0.001$ & $\pm 0.42$ & $\pm0.003$ & $\pm0.005$ \\ \cline{1-7}
 \hadr & $\pm 0.47$ & $\pm0.001$ & $\pm0.057$ & $\pm 0.48$ & $\pm0.001$ & $\pm0.065$ \\ \cline{1-7}
 \exptl & $\pm 0.69$ & $\pm0.004$ & $\pm0.010$ & $\pm 2.55$ & $\pm0.008$ & $\pm0.038$ \\ \cline{1-7}
 \tote & $\pm 1.17$ & $\pm0.006$ & $\pm0.060$ & $\pm 2.96$ & $\pm0.010$ & $\pm0.079$ \\ \cline{1-7}
 & \multicolumn{3}{c||}{\bt} \\ \cline{1-4}\cline{1-4}
 & \nf & \asmz & \anull \\ \cline{1-4}
 \stan &  3.77 & 0.108 & 0.635 \\ \cline{1-4}
 \fite & $\pm 1.12$ & $\pm0.003$ & $\pm0.019$ \\ \cline{1-4}
 \chisqd & \multicolumn{3}{c||}{ 132.1/137 } \\ \cline{1-4}
 \rensc & $\pm 0.46$ & $\pm0.004$ & $\pm0.006$ \\ \cline{1-4}
 \hadr & $\pm 0.33$ & $\pm0.001$ & $\pm0.089$ \\ \cline{1-4}
 \exptl & $\pm 2.97$ & $\pm0.009$ & $\pm0.002$ \\ \cline{1-4}
 \tote & $\pm 3.22$ & $\pm0.010$ & $\pm0.091$ \\ \cline{1-4}
\end{tabular}
\caption[ bla ]{
Results are shown for fits with the observables \thr, \cp\ and \bt\ to
\asmz, \anull\ and \nf. }
\label{tab_res_nf_a0free}
\end{center}
\end{table}

\begin{table}[!htb]
\begin{center}
\begin{tabular}{|r||c|c|c||c|c|c||} \hline
 & \multicolumn{3}{c||}{\thr} & \multicolumn{3}{c||}{\cp} \\ \cline{1-7}\cline{1-7}
 & \ca & \asmz & \anull & \ca & \asmz & \anull \\ \cline{1-7}
 \stan &  2.73 & 0.121 & 0.528 &  3.03 & 0.111 & 0.525 \\ \cline{1-7}
 \fite & $\pm 0.16$ & $\pm0.003$ & $\pm0.020$ & $\pm 0.25$ & $\pm0.004$ & $\pm0.029$ \\ \cline{1-7}
 \chisqd & \multicolumn{3}{c||}{ 128.1/235 } & \multicolumn{3}{c||}{ 103.4/154 } \\ \cline{1-7}
 \rensc & $\pm 0.05$ & $\pm0.004$ & $\pm0.001$ & $\pm 0.08$ & $\pm0.003$ & $\pm0.002$ \\ \cline{1-7}
 \hadr & $\pm 0.10$ & $\pm0.001$ & $\pm0.057$ & $\pm 0.09$ & $\pm0.001$ & $\pm0.064$ \\ \cline{1-7}
 \exptl & $\pm 0.14$ & $\pm0.004$ & $\pm0.012$ & $\pm 0.50$ & $\pm0.008$ & $\pm0.053$ \\ \cline{1-7}
 \tote & $\pm 0.24$ & $\pm0.006$ & $\pm0.061$ & $\pm 0.57$ & $\pm0.009$ & $\pm0.089$ \\ \cline{1-7}
 & \multicolumn{3}{c||}{\bt} \\ \cline{1-4}\cline{1-4}
 & \ca & \asmz & \anull \\ \cline{1-4}
 \stan &  3.24 & 0.108 & 0.626 \\ \cline{1-4}
 \fite & $\pm 0.21$ & $\pm0.003$ & $\pm0.023$ \\ \cline{1-4}
 \chisqd & \multicolumn{3}{c||}{ 131.8/137 } \\ \cline{1-4}
 \rensc & $\pm 0.07$ & $\pm0.004$ & $\pm0.006$ \\ \cline{1-4}
 \hadr & $\pm 0.06$ & $\pm0.001$ & $\pm0.088$ \\ \cline{1-4}
 \exptl & $\pm 0.60$ & $\pm0.009$ & $\pm0.086$ \\ \cline{1-4}
 \tote & $\pm 0.64$ & $\pm0.010$ & $\pm0.126$ \\ \cline{1-4}
\end{tabular}
\caption[ bla ]{
Results are shown for fits with the observables \thr, \cp\ and \bt\ to
\asmz, \anull\ and \ca. }
\label{tab_res_ca_a0free}
\end{center}
\end{table}

\begin{table}[!htb]
\begin{center}
\begin{tabular}{|r||c|c|c||c|c|c||} \hline
 & \multicolumn{3}{c||}{\thr} & \multicolumn{3}{c||}{\cp} \\ \cline{1-7}\cline{1-7}
 & \cf & \asmz & \anull & \cf & \asmz & \anull \\ \cline{1-7}
 \stan &  1.42 & 0.113 & 0.478 &  1.49 & 0.105 & 0.490 \\ \cline{1-7}
 \fite & $\pm 0.21$ & $\pm0.009$ & $\pm0.041$ & $\pm 0.31$ & $\pm0.011$ & $\pm0.074$ \\ \cline{1-7}
 \chisqd & \multicolumn{3}{c||}{ 130.4/235 } & \multicolumn{3}{c||}{ 103.2/154 } \\ \cline{1-7}
 \rensc & $\pm 0.03$ & $\pm0.006$ & $\pm0.013$ & $\pm 0.10$ & $\pm0.006$ & $\pm0.020$ \\ \cline{1-7}
 \hadr & $\pm 0.03$ & $\pm0.001$ & $\pm0.053$ & $\pm 0.06$ & $\pm0.002$ & $\pm0.059$ \\ \cline{1-7}
 \exptl & $\pm 0.21$ & $\pm0.011$ & $\pm0.022$ & $\pm 0.10$ & $\pm0.004$ & $\pm0.031$ \\ \cline{1-7}
 \tote & $\pm 0.30$ & $\pm0.015$ & $\pm0.072$ & $\pm 0.35$ & $\pm0.014$ & $\pm0.103$ \\ \cline{1-7}
 & \multicolumn{3}{c||}{\bt} & \multicolumn{3}{c||}{\bw} \\ \cline{1-7}\cline{1-7}
 & \cf & \asmz & \anull & \cf & \asmz & \anull \\ \cline{1-7}
 \stan &  2.28 & 0.085 & 0.406 &  3.50 & 0.052 & 0.311 \\ \cline{1-7}
 \fite & $\pm 0.47$ & $\pm0.009$ & $\pm0.080$ & $\pm 0.73$ & $\pm0.009$ & $\pm0.068$ \\ \cline{1-7}
 \chisqd & \multicolumn{3}{c||}{ 129.4/137 } & \multicolumn{3}{c||}{  79.1/120 } \\ \cline{1-7}
 \rensc & $\pm 0.88$ & $\pm0.029$ & $\pm0.222$ & $\pm 0.44$ & $\pm0.005$ & $\pm0.031$ \\ \cline{1-7}
 \hadr & $\pm 0.32$ & $\pm0.006$ & $\pm0.056$ & $\pm 0.38$ & $\pm0.004$ & $\pm0.051$ \\ \cline{1-7}
 \exptl & $\pm 0.80$ & $\pm0.018$ & $\pm0.248$ & $\pm 1.44$ & $\pm0.023$ & $\pm0.284$ \\ \cline{1-7}
 \tote & $\pm 1.32$ & $\pm0.036$ & $\pm0.348$ & $\pm 1.72$ & $\pm0.025$ & $\pm0.298$ \\ \cline{1-7}
\end{tabular}
\caption[ bla ]{
Results are shown for fits with the observables \thr, \cp, \bt\ and 
\bw\ to \asmz, \anull\ and \cf. }
\label{tab_res_cf_a0free}
\end{center}
\end{table}

\begin{table}[!htb]
\begin{center}
\begin{tabular}{|r||c|c|c|c||c|c|c|c||} \hline
 & \multicolumn{4}{c||}{\thr} \\ \cline{1-5}\cline{1-5}
 & \ca & \cf & \asmz & \anull \\ \cline{1-5}
 \stan &  2.68 &  1.21 & 0.128 & 0.567 \\ \cline{1-5}
 \fite & $\pm 0.19$ & $\pm 0.26$ & $\pm0.016$ & $\pm0.093$ \\ \cline{1-5}
 \chisqd & \multicolumn{4}{c||}{ 127.8/234 } \\ \cline{1-5}
 \rensc & $\pm 0.03$ & $\pm 0.02$ & $\pm0.005$ & $\pm0.006$ \\ \cline{1-5}
 \hadr & $\pm 0.16$ & $\pm 0.14$ & $\pm0.010$ & $\pm0.042$ \\ \cline{1-5}
 \exptl & $\pm 0.17$ & $\pm 0.35$ & $\pm0.025$ & $\pm0.164$ \\ \cline{1-5}
 \tote & $\pm 0.30$ & $\pm 0.46$ & $\pm0.032$ & $\pm0.194$ \\ \cline{1-5}
\end{tabular}
\caption[ bla ]{
Results are shown for fits with the observables \thr\ to \asmz, \anull,
\ca\ and \cf. }
\label{tab_res_cacf_a0free}
\end{center}
\end{table}

\begin{table}[!htb]
\begin{center}
\begin{tabular}{|l|r||c|c|c|c|}
\hline
fit type & & \anull & \nf & \ca & \cf \\
\hline\hline
\asmz-\anull     & \as & $-$0.87 &       &          &          \\ \hline
\asmz-\nf        & \as &         & 0.985 &          &          \\ \hline
\asmz-\ca        & \as &         &       & $-$0.981 &          \\ \hline
\asmz-\cf        & \as &         &       &          & $-$0.997 \\ \hline
\asmz-\ca-\cf    & \as &         &       &    0.20  & $-$0.68  \\
                 & \ca &         &       &          & $-$0.85  \\ \hline
\asmz-\anull-\nf & \as & 0.69    & 0.97  &          &          \\
                 & \anull &      & 0.82  &          &          \\ \hline
\asmz-\anull-\ca & \as & 0.78    &       & $-$0.97  &          \\
                 & \anull &      &       & $-$0.89  &          \\ \hline
\asmz-\anull-\cf & \as & 0.96    &       &          & $-$0.998 \\
                 & \anull &      &       &          & $-$0.975 \\ \hline
\asmz-\anull-\ca-\cf & \as & 0.988 &     & $-$0.72  & $-$0.985 \\
                 & \anull &      &       & $-$0.75  & $-$0.970 \\
                 & \ca &         &       &          & 0.59     \\
\hline
\end{tabular}
\caption[ bla ]{ The correlations between the fit parameters for
all types of standard fits to the \thr\ distributions are shown. }
\label{tab_corr}
\end{center}
\end{table}

\begin{table}
\begin{center}
\begin{tabular}{|l||c|c|c||c|c|c|c||}
\hline
 & \nf & \asmz & \anull & \ca & \asmz & \anull \\ \hline\hline
result & 5.64 & 0.116 & 0.524 & 2.88 & 0.116 & 0.526 \\ \hline
error  & $\pm1.35$ & $\pm0.005$ & $\pm0.064$ 
	& $\pm0.27$ & $\pm0.005$ & $\pm0.067$ \\ \hline\hline
 & \cf & \asmz & \anull & \ca & \cf & \asmz \\ \hline\hline
result & 1.45 & 0.109 & 0.484 & 2.84 & 1.29 & 0.119 \\ \hline
error & $\pm0.27$ & $\pm0.013$ & $\pm0.075$ 
      & $\pm0.24$ & $\pm0.18$ & $\pm0.10$  \\ \hline
\end{tabular}
\caption[bla]{ The combined results based on \thr\ and \cp\ are shown
with total errors. The results for \ca, \cf\ and \asmz\ are from fits
with \anull\ fixed. }
\label{tab_combres}
\end{center}
\end{table}

\clearpage

\section*{ Figures }

\end{document}